# Electronic Properties of Vinylene-Linked Heterocyclic Conducting Polymers: Predictive Design and Rational Guidance from DFT Calculations

*Bryan M. Wong\* and Joseph G. Cordaro*

Materials Chemistry Department, Sandia National Laboratories, Livermore, California 94551

\*Corresponding author. E-mail: bmwong@sandia.gov. Web: http://alum.mit.edu/www/usagi

The band structure and electronic properties in a series of vinylene-linked heterocyclic conducting polymers are investigated using density functional theory (DFT). In order to accurately calculate electronic band gaps, we utilize hybrid functionals with fully periodic boundary conditions to understand the effect of chemical functionalization on the electronic structure of these materials. The use of predictive first-principles calculations coupled with simple chemical arguments highlights the critical role that aromaticity plays in obtaining a low band gap polymer. Contrary to some approaches which erroneously attempt to lower the band gap by increasing the aromaticity of the polymer backbone, we show that being aromatic (or quinoidal) in itself *does not* insure a low band gap. Rather, an iterative approach which destabilizes the ground state of the parent polymer towards the aromatic ↔ quinoidal level-crossing on the potential energy surface is a more effective way of lowering the band gap in these conjugated systems. Our results highlight the use of predictive calculations guided by rational chemical intuition for designing low band gap polymers in photovoltaic materials.

**Keywords:** band structure, hybrid DFT, NICS, bond length alternation, aromatic, quinoidal

## 1. Introduction

Conducting polymers consisting of conjugated heterocyclic chains are one of the most frequently studied classes of organic materials due to their highly conjugated $\pi$-bonding systems, chemical stability, and tunable electronic properties. Although there has been significant progress in inorganic photovoltaic materials, much current research is now directed towards organic photovoltaics, which are potentially less expensive and easier to synthesize.[1-4] In particular, the relative ease in functionalizing organic materials using various electron donor/acceptor groups[5-9] allows the possibility of designing polymers with small band gaps *intrinsic* to the material itself, negating the need for further electrostatic doping of the system. Consequently, the increasing drive towards fully-organic systems has resulted in significant technological progress in next-generation organic field-effect transistors (OFETs),[10-12] organic light-emitting diodes (OLEDs),[13,14] and flexible photovoltaic materials.[15-17]

The conventional approach to developing novel conducting polymers is based on the chemical intuition of synthetic chemists which has had significant success in the past, but is ultimately time-consuming due to the nearly limitless number of promising candidate materials. Instead, the application of predictive computational design combined with chemical intuition leads to a rational and more efficient approach. From a practical point of view, the use of first-principles computational design to predict electronic properties beforehand offers significant advantages compared to the conventional approach: (1) it is inherently more time- and cost-efficient; and (2) it can dramatically reduce the number of promising synthetic targets for experimental design.[18-21] An important area where theory should specifically contribute is the prediction and rational understanding of how different chemical functional groups modulate electronic properties (and spectroscopic observables[22-25]) in order to ultimately guide the organic synthesis.[26-31]

In this work, we investigate the band structure and electronic properties in vinylene-linked heterocyclic polymers using density functional theory (DFT) calculations. We primarily focus on vinylene-linked polymers since experimental efforts in our group have shown that the vinylene group significantly enhances $\pi$-conjugation by delocalizing electrons along the backbone, leading to heterocyclic polymers with very low band gaps. Furthermore, vinylene groups offer a more flexible polymer chain, which could improve solubility compared to directly-coupled aryl-aryl conjugated polymers. Our DFT calculations utilize hybrid functionals with fully periodic boundary conditions (PBC) to obtain accurate band gaps in these conjugated polymers. It is important to mention that there has already been several studies on heterocyclic polymers and their electronic properties;[32-35] however, most of these studies primarily focused on isolated oligomers and did not address band-structure properties in a fully-periodic geometry. As a result, these oligomer calculations are only appropriate for molecular systems and do not capture the full electronic band structure as a function of electron momentum (i.e., molecular calculations on extended oligomers are incapable of determining whether a polymer has a direct (or indirect) band gap, which is an essential property for describing optoelectronic and electron-transport efficiencies in these systems). Indeed, the use of fully-periodic approaches for an accurate description of band structure and gap is mandatory since the modeling of systems using oligomers can introduce spurious border effects related to the finite size of the oligomer, potentially affecting the representation of the band structure (especially the conduction band[36,37]).

Following calculations on various vinylene-linked heterocyclic polymers, we then examine the effect of modifying the aromaticity of the polymer backbone via a benzannulation reaction or a change in the heterocyclic functional group itself. We highlight the importance of aromaticity in designing a low band gap polymer, and give rational guidance on designing a low band gap polymer using simple chemical arguments coupled with an iterative DFT evaluation of electronic properties. Our approach closely follows the analysis by Kertesz, et al.[38] and strongly emphasizes the somewhat non-intuitive idea that increasing the aromaticity of the polymer backbone (or even being aromatic in itself!) *does not* insure a small band gap. Rather, the destabilization of the ground state to shift molecular orbitals towards an aromatic ↔ quinoidal level-crossing on the potential energy surface will effectively decrease the band gap in these conjugated polymers. Finally, we show that the vinylene-linked, benzannulated pyrrole polymer is an interesting building block for further synthetic experiments, and we discuss the implications for rationally tuning the electronic properties in this conjugated system.

## 2. Computational Details

The semiconducting polymers analyzed in this work are shown in Figures 1(a) and 1(b). These particular polymers were chosen since they form a representative set of first- and second-row heterocycles which are also synthetically accessible. Our quantum chemical calculations consisted of two separate parts: (1) computation of electronic band structures and bond length alternation (BLA)[39-42] values using hybrid DFT with periodic boundary conditions and (2) evaluation of nucleus-independent chemical shift (NICS(1)) values[43] for optimized oligomers built from 9 monomer subunits. We describe each of these different types of calculations below.

**Periodic Boundary Calculations on Polymers.** All DFT calculations with periodic boundary calculations were carried out with the B3LYP hybrid functional which incorporates a fixed combination of 20% Hartree-Fock exchange and Becke's GGA correlation correction.[44] It is well-known that pure local and gradient-corrected functionals (i.e., LDA and PBE) severely underestimate semiconductor band gaps and do not account for excitonic effects,[45-47] while hybrid functionals partially overcome this problem by mixing in a fraction of nonlocal Hartree-Fock exchange.[48] We should also mention at this point that recent methodological progress has been made in using range-separated DFT techniques for both molecular applications[30,31,47,49-52] and periodic organic systems.[53-56] In particular, the recent Heyd-Scuseria-Ernzerhof (HSE) functional[56] incorporates a screened Hartree-Fock interaction for small distances and is, therefore, more computationally efficient than traditional hybrid functionals. We initially used the HSE functional (with the original recommended screening parameter of $\omega = 0.11\ a_0^{-1}$) to compute the electronic properties of our polymers, but found that the B3LYP functional demonstrated better agreement with experimental band gaps of polythiophene (experimental[57]: 2.1 eV, HSE: 1.68 eV, B3LYP: 2.04 eV, BHHLYP: 3.90 eV). The good agreement between B3LYP and experimental band gaps for semiconducting polymers has also been demonstrated very recently in the theoretical study of periodic organic polymers by Janesko.[58] Based on these studies, we have chosen the B3LYP functional since it also provides a balanced description of band gaps and excitonic effects in conjugated systems[45] (see Section 3.1, however, for some caveats of using B3LYP on calculating BLA and NICS(1) values). For this work, the infinite-chain polymer geometries and their unit cells were optimized at the B3LYP/6-31G(d,p) level of theory using one-dimensional periodic boundary conditions (reference Cartesian coordinates and total energies can be found in the Supporting Information). At the optimized periodic geometries, the BLA for each polymer was computed by taking the difference between the bond length of the (nominal) C–C linker bond attached to the heterocyclic ring and the adjacent vinylene linker group (i.e., BLA = $R_{C-C} - R_{vinylene}$). According to our chosen definition of the BLA, polymers with a small positive alternation show aromatic behavior while negative BLA values indicate quinoidal character. At the optimized geometries, a band structure calculation was performed with a larger 6-311G(d,p) basis set using 100 $k$ points along the Brillouin zone. In our single-point calculations with the large triple-zeta 6-311G(d,p) basis set, we found that the use of larger or more diffuse basis sets did not significantly improve electronic properties, especially when periodic boundary conditions were used.[59] All calculations were performed with Gaussian 09.[60]

**NICS(1) Calculations on Oligomers.** To give a quantitative measure of $\pi$-electron delocalization in these heterocyclic systems, we also calculated NICS(1) values to provide a relative comparison of aromaticity among all of the polymers. In the NICS(1) procedure suggested by Schleyer et al.[43] and extensively reviewed by Bachrach,[61] the absolute magnetic shielding is computed at 1 Å above and 1 Å below the geometric center of the ring (for the heterocyclic rings in this work, we define their center as the *nonweighted* mean of the heavy atom coordinates). In order to correspond with the NMR chemical shift convention, the sign of the absolute magnetic shielding is reversed to give the NICS(1) value. Similar to the BLA values, the resulting average NICS(1) values also give an electronic measure of $\pi$-orbital aromaticity, with more negative NICS(1) values denoting aromaticity, and more positive values corresponding to quinoidal character.[43,46,61] Since the available implementation of magnetic shieldings is restricted to molecules in Gaussian 09, we performed NICS(1) calculations on oligomeric sections for all of our polymers. A convergence study was done as a function of oligomer length, and we found that the NICS(1) values in the center of the oligomer did not change appreciably when the length reached 9 monomer subunits. We therefore took the NICS(1) values in the middle unit as being representative for the infinite polymer chain (we should also note that an extrapolation of the NICS(1) value as a function of the inverse number of unit cells, $1/n$, could also be used to accurately evaluate the infinite-length NICS(1) limit by using only a series of small oligomers, i.e., $n = 1 - 5$).[62] Geometries for all 18 nonamers were optimized at the B3LYP/6-31G(d,p) level of theory without symmetry constraints (reference Cartesian coordinates and total energies can be found in the Supporting Information). At the optimized geometries, NICS(1) values were calculated with a larger 6-311G(d,p) basis set for each of the rings within the middle segment of the nonamer.

## 3. Results and Discussion

### 3.1. Band gaps, BLA, and NICS(1) values

In each of the optimized infinite-chain polymers, the carbon atoms in the heterocycle become coplanar with the vinylene linking unit, indicating a strong conjugation across the valence $\pi$ orbitals of the C=C double bonds. In Table 1, we compare the band gaps, BLA, and NICS(1) values among the five-membered heterocyclic polymers, and Table 2 gives the corresponding information for the benzannulated polymers. As described in Section 2, the BLA values in both tables were evaluated from the optimized periodic geometries, while the NICS(1) values were calculated for rings in the middle of the isolated nonamer subunit. As a reference point in this work, the NICS(1) value for isolated benzene at the B3LYP/6-311G(d,p) level of theory is -11.1 ppm. It is important to mention at this point, however, a few caveats of using the B3LYP functional for calculating the band gaps, BLA, and NICS(1) values listed in Tables 1 and 2. While our previous comparison between B3LYP and experimental band gaps for polythiophene (Section 2) yielded exceptional accuracy, this direct comparison may be a fortuitous cancellation of several effects as it is well-known that B3LYP is less successful in describing other properties such as the BLA.[49,62-64] Indeed, for a few test cases on our five-membered heterocyclic polymers, we found that our B3LYP-derived BLA values (which incorporate 20% Hartree-Fock exchange) were underestimated with respect to BHHLYP calculations (defined with 50% Hartree-Fock exchange), in agreement with other previous studies.[62-64] Similarly, we also found that our B3LYP NICS(1) values were more positive than the BHHLYP ones, which is largely due to the different amounts of Hartree-Fock exchange in each functional.[62] In our benchmark cases, however, we found that the relative ordering of electronic properties for both B3LYP and BHHLYP were identical (i.e., there was no re-ordering of band gaps, BLA, or NICS(1) values between polymer species when using either functional), and that the overall trends in our B3LYP calculations were unaffected. A proper theoretical treatment to obtain both band gaps and accurate BLA values would require periodic-geometry optimizations and electronic properties at either the GW[65] or CCSD(T) level of theory (including possible effects such as molecular disorder and defects), which is beyond both the scope of the present paper and current computational technology. However, it is clear that our periodic B3LYP approach gives more realistic properties compared to widely-used LDA or GGA calculations, and our tabulated values are a reasonable choice for parametric studies on our large series of polymers.

To provide further insight into electronic properties, we also plot the band structure along the irreducible Brillouin zone (defined by the points $\Gamma$ and X) for each of the polymers. The electronic band structure gives a panoramic view of electronic energies and delocalization in each of the polymer systems. Specifically, the width of a particular electronic band reflects its orbital interactions along the polymer chain, with wide bands denoting delocalization and narrow bands corresponding to localization/small orbital overlap. For each of the different polymers, we found that their electronic band structures showed they were all semiconductors with a direct band gap at the Γ symmetry point (it should be noted, however, that not all one-dimensional polymers have direct band gaps; for example, several ladder-type polymers such as the fused polyborole structures in Ref. 66 have indirect band gaps). Figures 2(a) and 2(b) display the band structure for poly-pyrrole, poly-cyclopentadiene, and their benzannulated versions. Since all of our band structures show semiconducting behavior at 0 K, the Fermi energy in these semiconductors is by definition not unique,[67] and *any* energy in the gap can be chosen as the Fermi level (i.e., any energy in the gap separates occupied from unoccupied levels at 0 K). Therefore, we have arbitrarily chosen the Fermi energy level to lie at the top of the occupied valence bands in all of our figures. The band structures for all of the other polymers can be found in the Supporting Information.

### 3.2. Five-membered heterocyclic polymers

For the five-membered heterocyclic polymers in this work, Table 1 shows that the band gaps roughly correlate with the π-donor strengths (e.g. Hammett $\sigma$ values[68]) of the functional group. Specifically, the band gaps for these polymers increase in the following order: $CH_2$ < $PCH_3$ < Se < $SiH_2$, S, $PCH_3O$ < O < NH << $BCH_3$. Boron is significantly different than the other polymers since it has two electrons less per ring and is highly quinoid. From the tabulated BLA and NICS(1) values, we also note that there is a noticeable correlation (though it is not perfect) between the band gap and aromaticity of the heterocycle. With the exception of boron, the energy gaps are smaller for the polymers built from non-aromatic heterocycles. For these polymers, the lowering of the band gap results from a stabilization of the lowest unoccupied crystal orbital (LUCO) compared to their aromatic counterparts (cf. band structures for NH vs. $CH_2$).

### 3.3. Benzannulated polymers

Next, we consider the electronic properties of the benzannulated polymers listed in Table 2. In comparing the results of the previous section, we note a drastic re-ordering of energy gaps and electronic properties. For the benzannulated polymers, the band gaps increase in the following order: NH << O, S < Se < PCH$_3$, BCH$_3$ < CH$_2$ < SiH$_2$ < PCH$_3$O. In particular, we find that the band gap of benzannulated poly-pyrrole (NH functional group) is considerably reduced compared to its non-benzannulated form. It is interesting to note that all of the benzannulated polymers with large band gaps also have band structures with very narrow LUCO bandwidths (cf figures in the Supporting Information). In other words, since the LUCO bandwidth is small, the conduction orbitals are highly localized (small orbital overlap), and electron mobility in these polymers will also be very low. From these results, some questions naturally arise: Why does the fusion of a benzene ring result in such a drastic band-gap reordering for the benzannulated polymers? Can we understand these effects to design a low-band-gap polymer using a combination of chemical intuition and first-principles calculations? In order to answer these questions, we must take a closer look at the variations in aromatic character within the polymers.

First, in each of the non-benzannulated heterocyclic polymers (regardless of the type of heteroatom), an electronic competition exists between maintaining the aromaticity of the individual heterocyclic rings versus delocalization along the backbone chain. It is also important to realize that the ground state of a given polymer is *not* always the aromatic configuration; i.e., the lowest-energy ground state may actually be quinoid, depending on the type of heteroatom. For example, the ground state of poly-pyrrole is highly aromatic, as demonstrated by its large, negative NICS(1) value in Table 1. Disrupting the aromaticity of this heterocycle causes a transition through a level-crossing to a quinoid form which lies higher in energy. Figure 3(a) shows a schematic of the highest occupied crystal orbitals (HOCOs) and LUCOs connecting the aromatic and quinoid forms through a level-crossing for poly-pyrrole (notice that all double and single bonds in the aromatic form transform into single and double bonds respectively in the quinoid structure, which is consistent with our definition of the BLA in Section 2). In contrast, the ground state of poly-cyclopentadiene ($CH_2$ functional group) is *quinoidal*, with a NICS(1) value of -3.0 ppm. As such, disrupting the quinoidal structure of poly-cyclopentadiene creates an aromatic form which actually lies *higher* in energy, as shown in Figure 3(b) (for both Figures 3(a) and 3(b), we arbitrarily define the reaction coordinate to point in the direction of increasing quinoid character. For example, if the BLA is chosen as the reaction coordinate, its zero-point reference delineates the aromatic/quinoidal boundary with small positive BLAs showing aromatic behavior and negative BLA values indicating quinoidal character). In both the poly-pyrrole and poly-cyclopentadiene examples, destabilizing the ground state creates a new polymer with a lower band gap: the HOCO of the new polymer is raised in energy due to this destabilization, and the LUCO of the new polymer is lowered since it has the same orbital configuration (cf. orbital lobes in Figures 3(a) and 3(b)) as the ground-state HOCO in the original stable polymer.

When a five-membered polymer is benzannulated, the resulting band gap will either increase or decrease depending on the aromaticity of the original parent polymer. During the benzannulation, a new electronic competition exists between maintaining the aromaticity of the fused benzene ring versus that of the heterocycle. Since the benzene ring (usually) has a larger resonance energy than the heterocycle, the heterocycle will adopt additional quinoid character to maintain the aromaticity of benzene. These facts coupled with first-principles calculations can be used as a qualitative guideline for designing new conjugated polymers with small intrinsic band gaps. Returning to our example of poly-pyrrole, we recall that the ground state of this polymer is highly aromatic. Fusing a benzene ring onto pyrrole will destabilize the parent polymer by increasing the quinoidal character of the pyrrole ring. Consequently, the benzannulation procedure shifts the orbitals towards a level-crossing which simultaneously destabilizes the HOCO and stabilizes the LUCO to yield a reduced band gap as depicted in Figure 4(a) (notice the similarities in crystal orbitals between Figures 3(a) and 4(a)). In contrast, the ground state of the poly-cyclopentadiene parent polymer is the quinoidal form. Fusing a benzene ring onto cyclopentadiene will still increase the quinoidal character of cyclopentadiene; however, since the ground state of the parent polymer *already* favors the quinoidal structure, the benzannulation procedure shifts the orbitals *away* from the level-crossing to further stabilize the parent polymer (notice the difference in crystal orbitals between Figures 3(b) and 4(b)). As a result, the HOCO is stabilized even further, and the LUCO is destabilized to yield a very large band gap. A similar rearrangement of band gaps is also observed for the other five-membered heterocyclic polymers as they are benzannulated. Comparing the band gaps listed in Tables 1 and 2, heterocycles which are aromatic (i.e., NH, O, S, and Se) undergo a band-gap lowering in their benzannulated forms. In contrast, heterocycles (with the exception of boron) which are quinoidal in the ground state yield large band gaps under benzannulation.

Lastly, to emphasize that destabilization of the ground state (i.e. moving towards a level-crossing) is key to designing a low-band-gap polymer and that being aromatic or quinoidal *in itself* does not insure a small band gap, we extend our previous example of benzannulated poly-pyrrole. As mentioned previously, the ground state of poly-pyrrole is aromatic; after benzannulation, a new low-band-gap polymer is formed whose ground state favors a quinoidal pyrrole ring. In order to further reduce the band gap of benzannulated poly-pyrrole, it is tempting to fuse additional benzene rings to increase the quinoidal character of the pyrrole ring even further. However, this procedure actually increases the band gap instead of lowering it, as demonstrated by the periodic B3LYP/6-311G(d,p) calculations listed in Figure 5. This evolution of band gap energies as a function of increasing quinoidal character in poly-pyrrole is also completely consistent with the overall trends in our other polymer species. In Figure 6, we plot the band gaps for both the five-membered and benzannulated heterocyclic polymers as a function of the BLA (note that in order to maintain consistency between our definitions of the BLA [Section 2] and the direction of the reaction coordinate [Section 3.3], the reaction coordinate in Figure 6 must point towards the left). The BLA = 0 reference point delineates the aromatic/quinoidal boundary with small positive BLAs showing aromatic behavior and negative BLA values indicating quinoidal character. As can be easily seen in Figure 6, being aromatic or quinoidal in itself (as indicated by the sign of the BLA) does not insure a low band gap. In fact, further increasing the aromatic or quinoidal character of the polymer by moving away from the BLA = 0 reference point (i.e. via repeated benzannulation of pyrrole) will only increase the band gap. Based on these simple trends between aromatic and quinoidal properties, we can construct the general band gap diagram shown in Figure 7 (note the overall similarity with Figure 6). In this very general schematic, the band gap in the lower panel is plotted continuously as a function of a generalized reaction coordinate, whose orientation we have chosen to point in the direction of increasing quinoid character (consistent with Figures 3 – 4). The band gap of the polymer attains its lowest value when the reaction coordinate is at the level-crossing/avoided-crossing denoted by $r_{opt}$, which separates aromatic from quinoidal character. Returning to our example of poly-pyrrole, the ground state of this (non-benzannulated) aromatic polymer lies to the far left of $r_{opt}$ in the band gap diagram. The first benzannulation of pyrrole increases quinoidal character and shifts the system closer to $r_{opt}$, effectively lowering the band gap. Fusing additional benzene rings will increase quinoidal character even more and move the system further to the right of $r_{opt}$, resulting in larger band gaps with each additional benzene ring. These results, in conjunction with Figures 6 – 7, clearly emphasize that the relative distance from the level-crossing, $r - r_{opt}$ directly affects the band gap, and not its sign (the sign of $r - r_{opt}$ indicates absolute aromatic/quinoidal character only). In order to reduce the band gap of benzannulated poly-pyrrole, one must destabilize the ground state to shift its orbitals back toward the level-crossing. Since the ground state of benzannulated poly-pyrrole is the quinoidal form, the band gap can be lowered by *increasing* its aromatic character in the next chemical-functionalization step.

**Conclusion**

In this study, we have investigated the band structure and electronic properties in a series of vinylene-linked organic polymers for photovoltaic applications. To understand and accurately predict the electronic properties in these materials, we utilized hybrid functionals with fully periodic boundary conditions (PBC) in conjunction with BLA and NICS(1) calculations to rationalize the critical role that aromaticity plays in these polymers. The use of predictive first-principles calculations coupled with chemical intuition is a promising route to designing a low band gap polymer, avoiding time-consuming experimental efforts that could lead to inefficient materials. More specifically, our results strongly emphasize that increasing the aromaticity of the polymer backbone (an erroneous approach which is sometimes followed in synthetic attempts) does not insure a low band gap. Instead, the destabilization of the ground state in the parent polymer towards the aromatic ↔ quinoidal level-crossing is a more effective way of lowering the band gap in these conjugated systems. Finally, based on our discussion of level-crossings and their effect on band gaps, we also draw attention towards the vinylene-linked, benzannulated pyrrole polymer as a low band gap material with very interesting electronic properties. Further tuning of the HOCO and LUCO energy levels to ensure air-oxidation strength of the pyrrole system would offer additional promising opportunities in low band gap polymer synthesis. Our computational screening methodology, coupled with chemical intuition on competing aromatic/quinoidal effects, allows a rational approach to designing low band gap semiconducting polymers for guided experimental efforts.

**Acknowledgement.** This research was supported in part by the National Science Foundation through TeraGrid resources (Grant No. TG-CHE1000066N) provided by the National Center for Supercomputing Applications. Funding for this effort was provided by the Readiness in Technical Base and Facilities (RTBF) program at Sandia National Laboratories, a multiprogram laboratory operated by Sandia Corporation, a Lockheed Martin Company, for the United States Department of Energy's National Nuclear Security Administration under contract DE-AC04-94AL85000.

**Supporting Information Available:** Electronic band structures and Cartesian coordinates for the 5-membered $BCH_3$, $CH_2$, $NH$, $O$, $PCH_3$, $PCH_3O$, $SiH_2$, $S$, and $Se$ heterocyclic polymers/nonamers and their benzannulated versions. This material is available free of charge via the Internet at http://pubs.acs.org.

**References and Notes**

(1) Lane, P. A.; Kafafi, Z. H. Solid-state organic photovoltaics: a review of molecular and polymeric devices. In *Organic Photovoltaics: Mechanisms, Materials, and Devices*; Sun, S., Sariciftci, N. S., Eds.; CRC Press: Boca Raton, FL, 2005; pp 49-104.

(2) Picciolo, L. C.; Murata, H.; Kafafi, Z. H. *Appl. Phys. Lett.* **2001**, *78*, 2378-2380.

(3) Kelley, T. W.; Baude, P. F.; Gerlach, C.; Ender, D. E.; Muyres, D.; Haase, M. A.; Vogel, D. E.; Theiss, S. D. *Chem. Mater*. **2004**, *16*, 4413-4422.

(4) Zade, S. S.; Bendikov, M. *Angew. Chem. Int. Ed.* **2010**, *49*, 4012-4015.

(5) Dimitrakopoulos, C. D.; Malenfant, P. R. L. *Adv. Mater.* **2002**, *14*, 99-117.

(6) Odom, S. A.; Parkin, S. R.; Anthony, J. E. *Org. Lett.* **2003**, *5*, 4245-4248.

(7) Rodriguez, M. A.; Zifer, T.; Vance, A. L.; Wong, B. M.; Leonard, F. *Acta Cryst.* **2008**, *E64*, o595.

(8) Rodriguez, M. A.; Nichol, J. L.; Zifer, T.; Vance, A. L.; Wong, B. M.; Léonard, F. *Acta Cryst.* **2008**, *E64*, o2258.

(9) Anthony, J. E. *Angew. Chem. Int. Ed.* **2008**, *47*, 452-483.

(10) Simmons, J. M.; In, I.; Campbell, V. E.; Mark, T. J.; Léonard, F.; Gopalan, P.; Eriksson, M. A. *Phys. Rev. Lett.* **2007**, *98*, 086802.

(11) Zhou, X.; Zifer, T.; Wong, B. M.; Krafcik, K. L.; Léonard, F.; Vance, A. L. *Nano Lett.* **2009**, *9*, 1028-1033.

(12) Wong, B. M.; Morales, A. M. *J. Phys. D: Appl. Phys.* **2009**, *42*, 055111.

(13) Pardo, D. A.; Jabbour, G. E.; Peyghambarian, N. *Adv. Mater.* **2000**, *12*, 1249-1252.

(14) Yang, X.; Neher, D.; Hertel, D.; Däubler, T. K. *Adv. Mater.* **2004**, *16*, 161-166.

(15) Davis, D. A.; Hamilton, A.; Yang, J.; Cremar, L. D.; Gough, D. V.; Potisek, S. L.; Ong, M. T.; Braun, P. V.; Martinez, T. J.; White, S. R.; Moore, J. S.; Sottos, N. R. *Nature* **2009**, *459*, 68-72.

(16) Potisek, S. L.; Davis, D. A.; Sottos, N. R.; White, S. R.; Moore, J. S. *J. Am. Chem. Soc.* **2007**, *129*, 13808-13809.

(17) O'Bryan, G.; Wong, B. M.; McElhanon, J. R. *ACS Appl. Mater. Interfaces* **2010**, *2*, 1594-1600.

(18) Zhou, X. W.; Zimmerman, J. A.; Wong, B. M.; Hoyt, J. J. *J. Mater. Res.* **2008**, *23*, 704-718.

(19) Wong, B. M.; Lacina, D.; Nielsen, I. M. B.; Graetz, J.; Allendorf, M. D. *J. Phys. Chem. C* **2011**, *115*, 7778-7786.

(20) Ward, D. K.; Zhou, X. W.; Wong, B. M.; Doty, F. P.; Zimmerman, J. A. *J. Chem. Phys.* **2011**, *134*, 244703.

(21) Wong, B. M.; Léonard, F.; Li, Q.; Wang, G. T. *Nano Lett.* **2011**, in press.

(22) Wong, B. M.; Steeves, A. H.; Field, R. W. *J. Phys. Chem. B* **2006**, *110*, 18912-18920.

(23) Bechtel, H. A.; Steeves, A. H.; Wong, B. M.; Field, R. W. *Angew. Chem. Int. Ed.* **2008**, *47*, 2969-2972.

(24) Wong, B. M. *Phys. Chem. Chem. Phys.* **2008**, *10*, 5599-5606.

(25) Wong, B. M. *J. Comput. Chem.* **2009**, *30*, 51-56.

(26) Dey, K. R.; Wong, B. M.; Hossain, M. A. *Tetrahedron Lett.* **2010**, *51*, 1329-1332.

(27) Hossain, M. A.; Saeed, M. A.; Fronczek, F. R.; Wong, B. M.; Dey, K. R. Mendy, J. S.; Gibson, D. *Cryst. Growth Des.* **2010**, *10*, 1478-1481.

(28) Saeed, M. A.; Wong, B. M.; Fronczek, F. R.; Venkatraman, R.; Hossain, M. A. *Cryst. Growth Des.* **2010**, *10*, 1486-1488.

(29) Işiklan, M.; Saeed, M. A.; Pramanik, A.; Wong, B. M.; Fronczek, F. R.; Hossain, M. A. *Cryst. Growth Des.* **2011**, *11*, 959-963.

(30) Wong, B. M.; Cordaro, J. G. *J. Chem. Phys.* **2008**, *129*, 214703.

(31) Wong, B. M.; Piacenza, M.; Della Sala, F. *Phys. Chem. Chem. Phys.* **2009**, *11*, 4498-4508.

(32) Hong, S. Y.; Kwon, S. J.; Kim, S. C. *J. Chem. Phys.* **1995**, *103*, 1871-1877.

(33) Salzner, U.; Lagowski, J. B.; Pickup, P. G.; Poirier, R. A. *Synth. Met.* **1998**, *96*, 177-189.

(34) Yamaguchi, S.; Itami, Y.; Tamao, K. *Organometallics* **1998**, *17*, 4910-4916.

(35) Delaere, D.; Nguyen, M. T.; Vanquickenborne, L. G. *Phys. Chem. Chem. Phys.* **2002**, *4*, 1522-1530.

(36) Deák, P. *Phys. Status Solidi B* **2000**, *217*, 9-21.

(37) Zunger, A. *J. Phys. C* **1974**, *7*, 76-96.

(38) Kertesz, M.; Choi, C. H.; Yang, S. *Chem. Rev.* **2005**, *105*, 3448-3481.

(39) Brédas, J. L. *J. Chem. Phys.* **1985**, *82*, 3808-3811.

(40) Brédas, J. L.; Heeger, A. J.; Wudl, F. *J. Chem. Phys.* **1986**, *85*, 4673-4678.

(41) Toussaint, J. M.; Wudl, F.; Brédas, J. L. *J. Chem. Phys.* **1989**, *91*, 1783-1788.

(42) Toussaint, J. M.; Brédas, J. L. *J. Chem. Phys.* **1991**, *94*, 8122-8128.

(43) Schleyer, P. v. R.; Manoharan, M.; Wang, Z.-X.; Kiran, B.; Jiao, H.; Puchta, R.; Hommes, N. J. R. v. E. *Org. Lett.* **2001**, *3*, 2465-2468.

(44) Becke, A. D. *J. Chem. Phys.* **1993**, *98*, 5648-5652.

(45) Tretiak, S.; Igumenshchev, K.; Chernyak, V. *Phys. Rev. B* **2005**, *71*, 033201.

(46) Wong, B. M. *J. Phys. Chem. C* **2009**, *113*, 21921-21927.

(47) Wong, B. M.; Hsieh, T. H. *J. Chem. Theory Comput.* **2010**, *6*, 3704-3712.

(48) Igumenshchev, K. I.; Tretiak, S.; Chernyak, V. Y. *J. Chem. Phys.* **2007**, *127*, 114902.

(49) Jacquemin, D.; Perpète, E. A.; Scalmani, G.; Frisch, M. J.; Kobayashi, R.; Adamo, C. *J. Chem. Phys.* **2007**, *126*, 144105.

(50) Jacquemin, D.; Perpète, E. A.; Vydrov, O. A.; Scuseria, G. E.; Adamo, C. *J. Chem. Phys.* **2007**, *127* 094102.

(51) Jacquemin, D.; Perpète, E. A.; Scuseria, G. E.; Ciofini, I.; Adamo, C. *J. Chem. Theory Comput.* **2008**, *4*, 123-135.

(52) Kornobis, K.; Kumar, N.; Wong, B. M.; Lodowski, P.; Jaworska, M.; Andruniów, T.; Ruud, K.; Kozlowski, P. M. *J. Phys. Chem. A* **2011**, *115*, 1280-1292.

(53) Heyd, J.; Scuseria, G. E.; Ernzerhof, M. *J. Chem. Phys.* **2003**, *118*, 8207-8215.

(54) Heyd, J.; Scuseria, G. E. *J. Chem. Phys.* **2004**, *120*, 7274-7280.

(55) Heyd, J.; Peralta, J. E.; Scuseria, G. E.; Martin, R. L. *J. Chem. Phys.* **2005**, *123*, 174101.

(56) Heyd, J.; Scuseria, G. E.; Ernzerhof, M. *J. Chem. Phys.* **2006**, *124*, 219906.

(57) Tsumura, A.; Koezuka, H.; Ando, T. *Appl. Phys. Lett.* **1986**, *49*, 1210-1212.

(58) Janesko, B. G. *J. Chem. Phys.* **2011**, *134*, 184105.

(59) Wong, B. M.; Ye, S. H. *Phys. Rev. B* **2011**, in press.

(60) Gaussian 09, Revision A.2, Frisch, M. J.; Trucks, G. W.; Schlegel, H. B.; Scuseria, G. E.; Robb, M. A.; Cheeseman, J. R.; Scalmani, G.; Barone, V.; Mennucci, B.; Petersson, G. A.; et al. Gaussian, Inc., Wallingford CT, 2009.

(61) Bachrach, S. M. Nucleus-Independent Chemical Shift (NICS). In *Computational Organic Chemistry*; John Wiley & Sons: Hoboken, NJ, **2007**; pp. 81-86.

(62) Peach, M. J. G.; Tellgren, E. I; Salek, P.; Helgaker, T.; Tozer, D. J. *J. Phys. Chem. A* **2007**, *111*, 11930-11935.

(63) Zhao, Y.; Truhlar, D. G. *J. Phys. Chem. A* **2006**, *110*, 10478-10486.

(64) Jacquemin, D.; Adamo, C. *J. Chem. Theory Comput.* **2011**, *7*, 369-376.

(65) van der Horst, J.-W.; Bobbert, P. A.; Michels, M. A. J.; Brocks, G.; Kelly, P. J. *Phys. Rev. Lett.* **1999**, *83*, 4413.

(66) Pesant, S.; Dumont, G.; Langevin, S.; Côté, M. *J. Chem. Phys.* **2009**, *130*, 114906.

(67) Ashcroft, N. W.; Mermin, N. D. In *Solid State Physics*; Sounders College Publishing: Philadelphia, PA, **1976**.

(68) Hammett, L. P. *Chem. Rev.* **1935**, *17*, 125-136.

**TABLE 1**: Band gaps, bond length alternations, and NICS(1) values for vinylene-linked heterocyclic polymers.

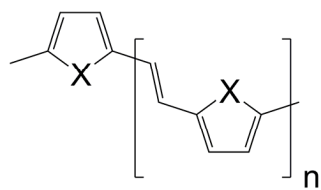

| (functional group X) | Band Gap (eV)[a] | Bond Length Alternation (Å)[b] | Heterocyclic Ring NICS(1) (ppm)[c] |
|---|---|---|---|
| BCH$_3$ | 2.38 | -0.059 | 0.0 |
| CH$_2$ | 1.12 | 0.054 | -3.0 |
| NH | 1.85 | 0.073 | -8.2 |
| O | 1.76 | 0.066 | -7.6 |
| PCH$_3$ | 1.22 | -0.050 | -3.9 |
| PCH$_3$O | 1.68 | -0.066 | -1.9 |
| SiH$_2$ | 1.65 | -0.061 | -0.8 |
| S | 1.68 | 0.074 | -7.5 |
| Se | 1.57 | 0.069 | -6.6 |

[a]Computed from periodic B3LYP/6-311G(d,p) single-point energy calculations on periodic B3LYP/6-31G(d,p)-optimized geometries

[b]Difference in bond length between the C–C bond and the adjacent vinylene linker group ($R_{C-C} - R_{vinylene}$); computed from periodic B3LYP/6-31G(d,p)-optimized geometries

[c]Computed from B3LYP/6-311G(d,p) NICS(1) calculations on B3LYP/6-31G(d,p)-optimized geometries for an isolated nonamer subunit. As a reference point in this work, the NICS(1) value for isolated benzene at the B3LYP/6-311G(d,p) level of theory is -11.1 ppm

**TABLE 2**: Band gaps, bond length alternations, and NICS(1) values for vinylene-linked, benzannulated heterocyclic polymers.

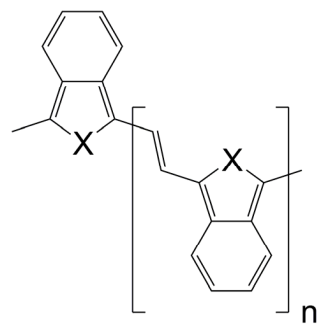

| (functional group X) | Band Gap (eV)[a] | Bond Length Alternation (Å)[b] | Heterocyclic Ring NICS(1) (ppm)[c] | Benzene Ring NICS(1) (ppm)[c] |
|---|---|---|---|---|
| BCH$_3$ | 2.30 | -0.062 | 0.1 | -7.9 |
| CH$_2$ | 2.44 | -0.077 | -2.2 | -8.7 |
| NH | 0.82 | -0.036 | -8.9 | -8.9 |
| O | 1.47 | -0.061 | -5.3 | -8.9 |
| PCH$_3$ | 2.28 | -0.072 | -2.2 | -8.7 |
| PCH$_3$O | 2.76 | -0.082 | -1.9 | -8.7 |
| SiH$_2$ | 2.64 | -0.077 | -1.3 | -8.4 |
| S | 1.49 | -0.056 | -4.7 | -9.2 |
| Se | 1.60 | -0.061 | -3.9 | -9.2 |

[a]Computed from periodic B3LYP/6-311G(d,p) single-point energy calculations on periodic B3LYP/6-31G(d,p)-optimized geometries

[b]Difference in bond length between the C–C bond and the adjacent vinylene linker group ($R_{C-C} - R_{vinylene}$); computed from periodic B3LYP/6-31G(d,p)-optimized geometries

[c]Computed from B3LYP/6-311G(d,p) NICS(1) calculations on B3LYP/6-31G(d,p)-optimized geometries for an isolated nonamer subunit. As a reference point in this work, the NICS(1) value for isolated benzene at the B3LYP/6-311G(d,p) level of theory is -11.1 ppm

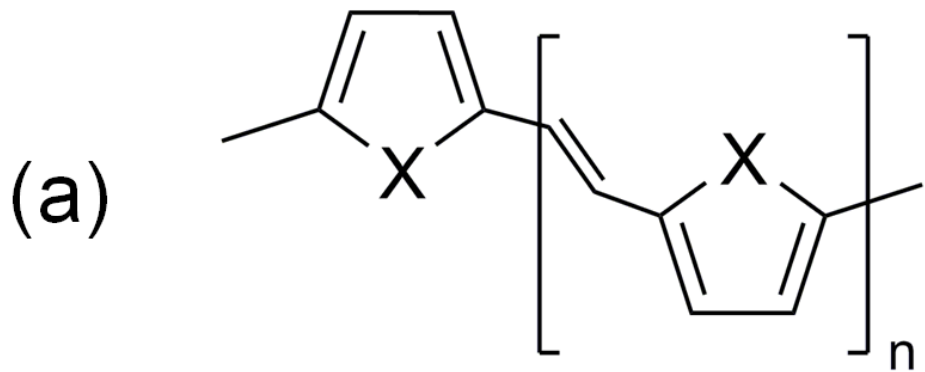

(a)

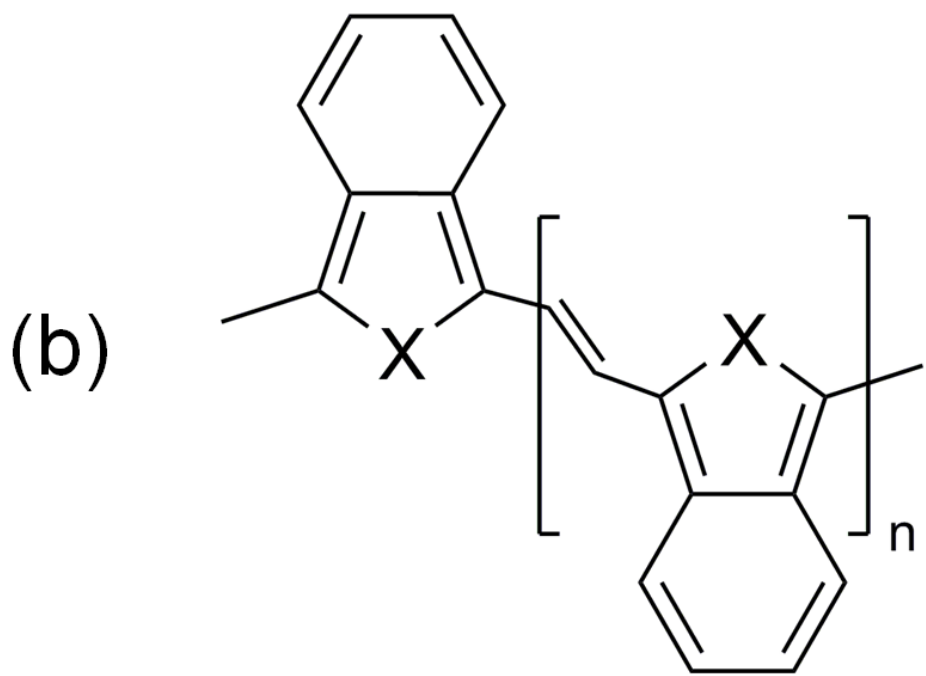

(b)

X = BCH₃, CH₂, NH, O, PCH₃, PCH₃O, SiH₂, S, Se

**Figure 1**

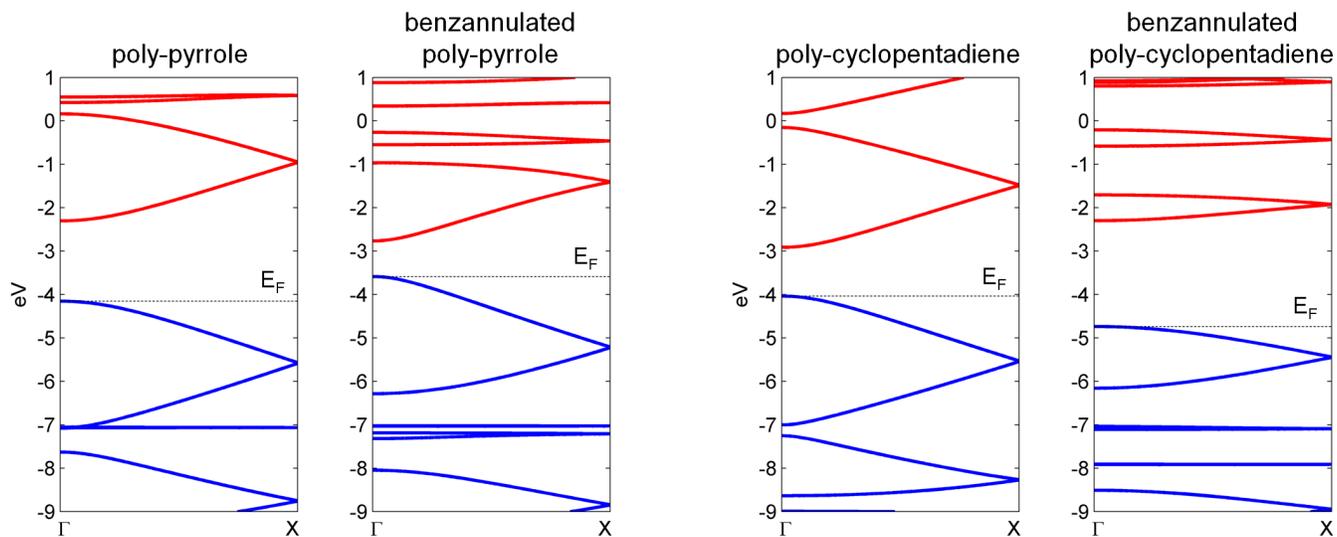

**Figure 2**

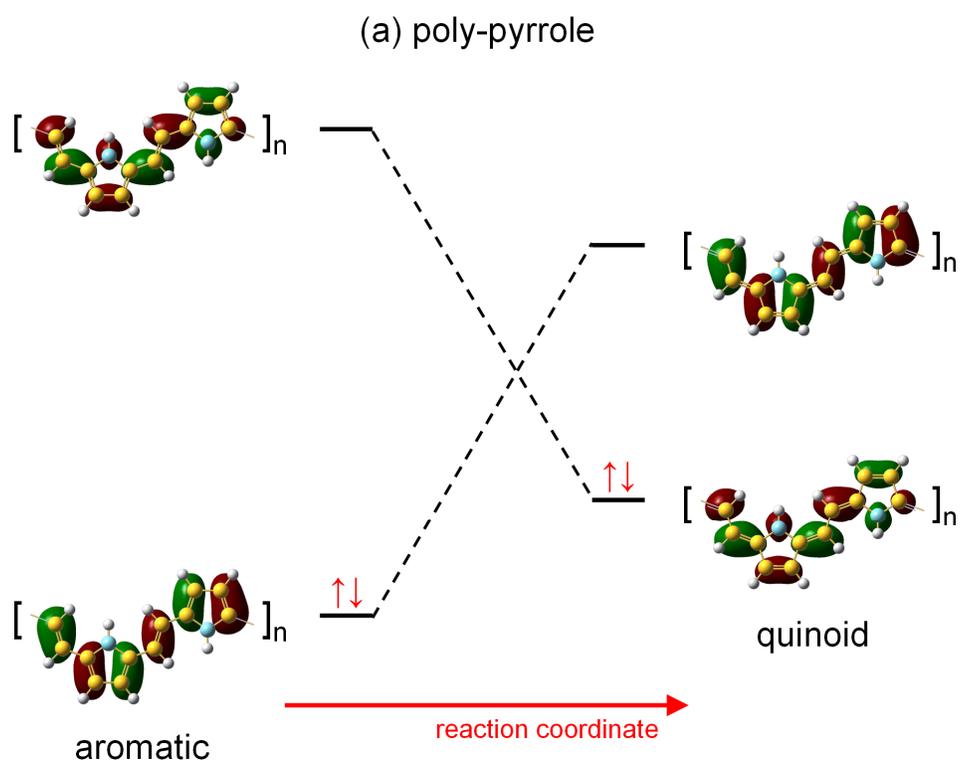

(a) poly-pyrrole

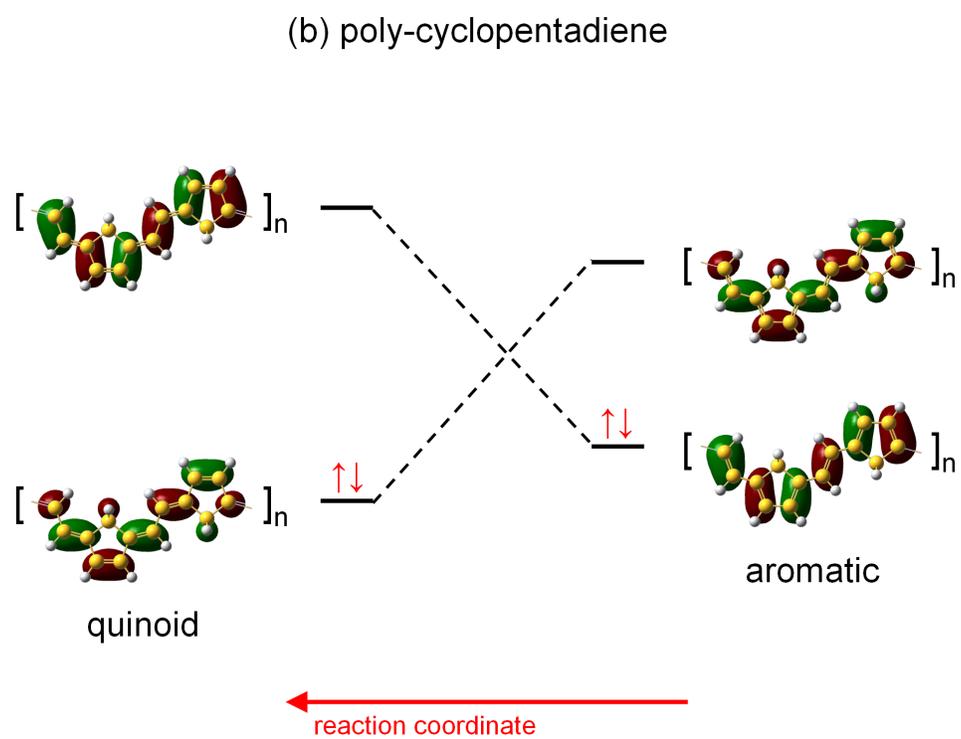

(b) poly-cyclopentadiene

**Figure 3**

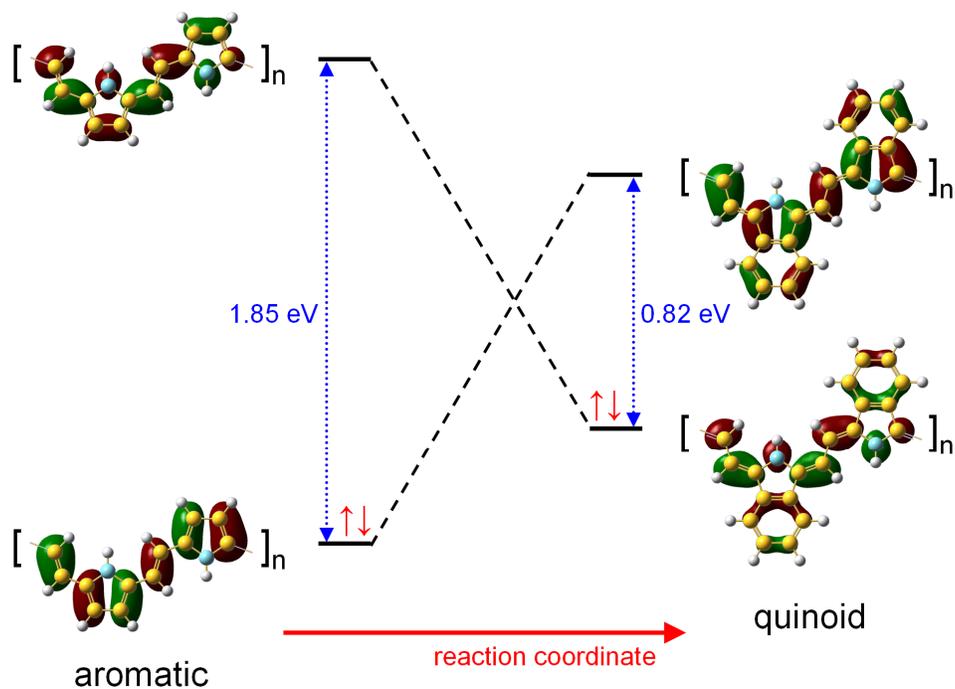

(a) benzannulation of poly-pyrrole

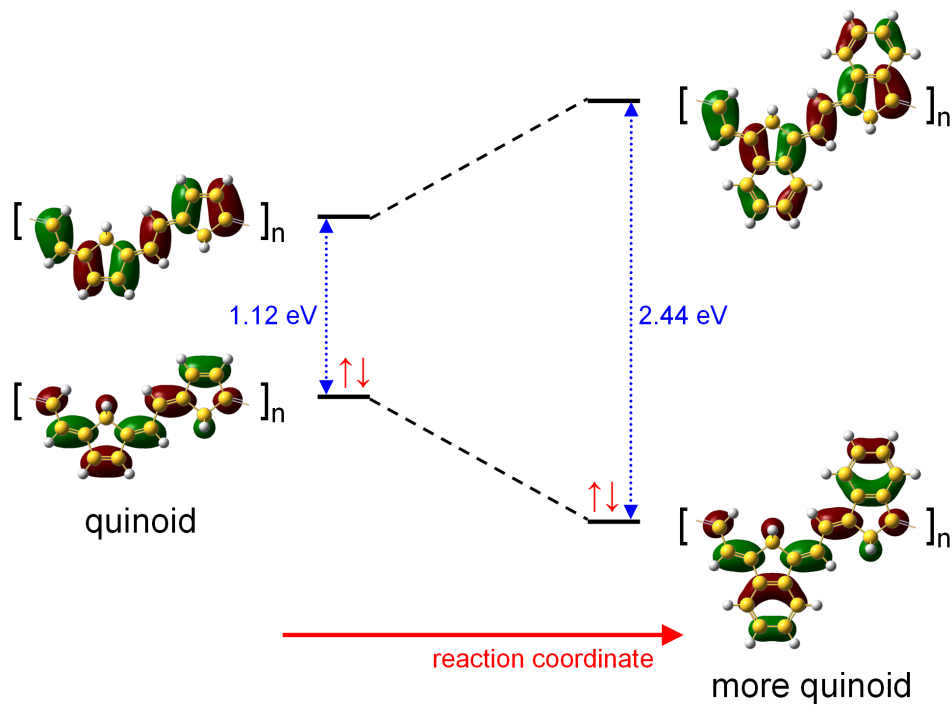

(b) benzannulation of poly-cyclopentadiene

**Figure 4**

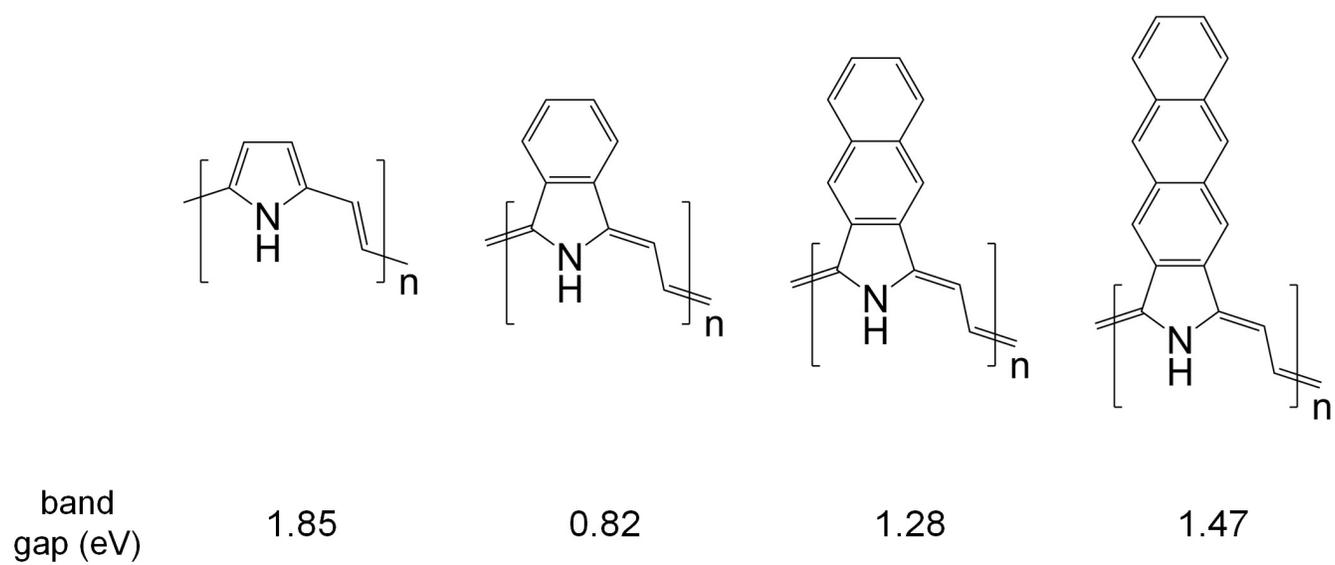

**Figure 5**

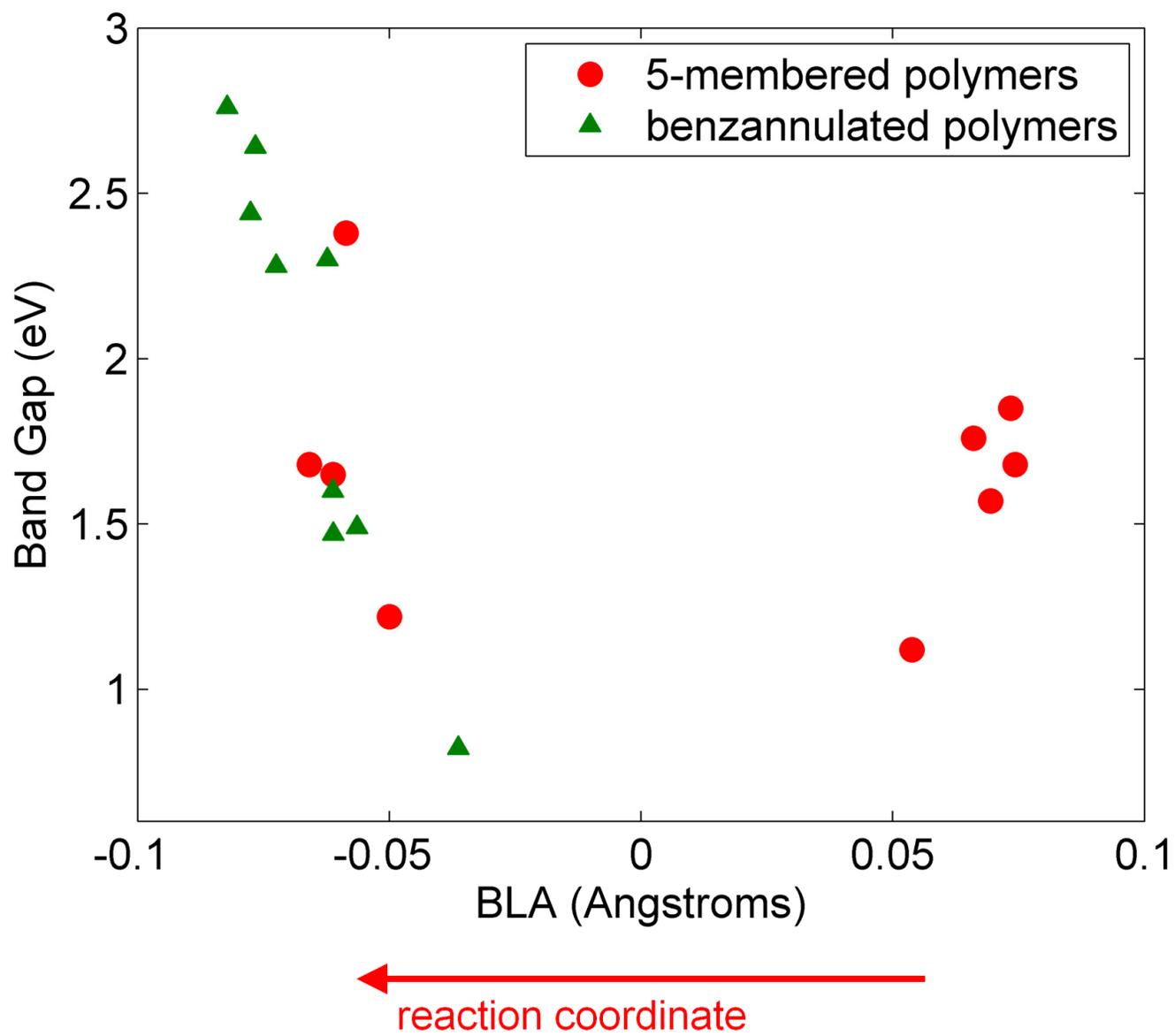

**Figure 6**

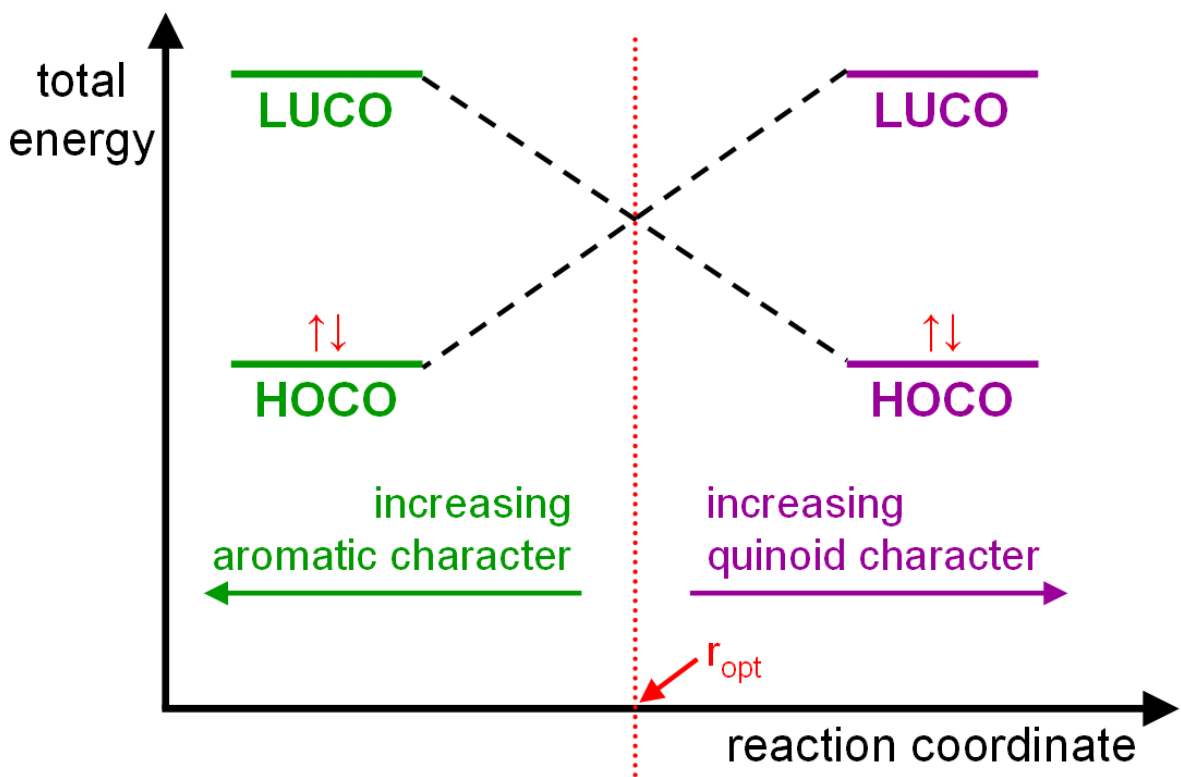

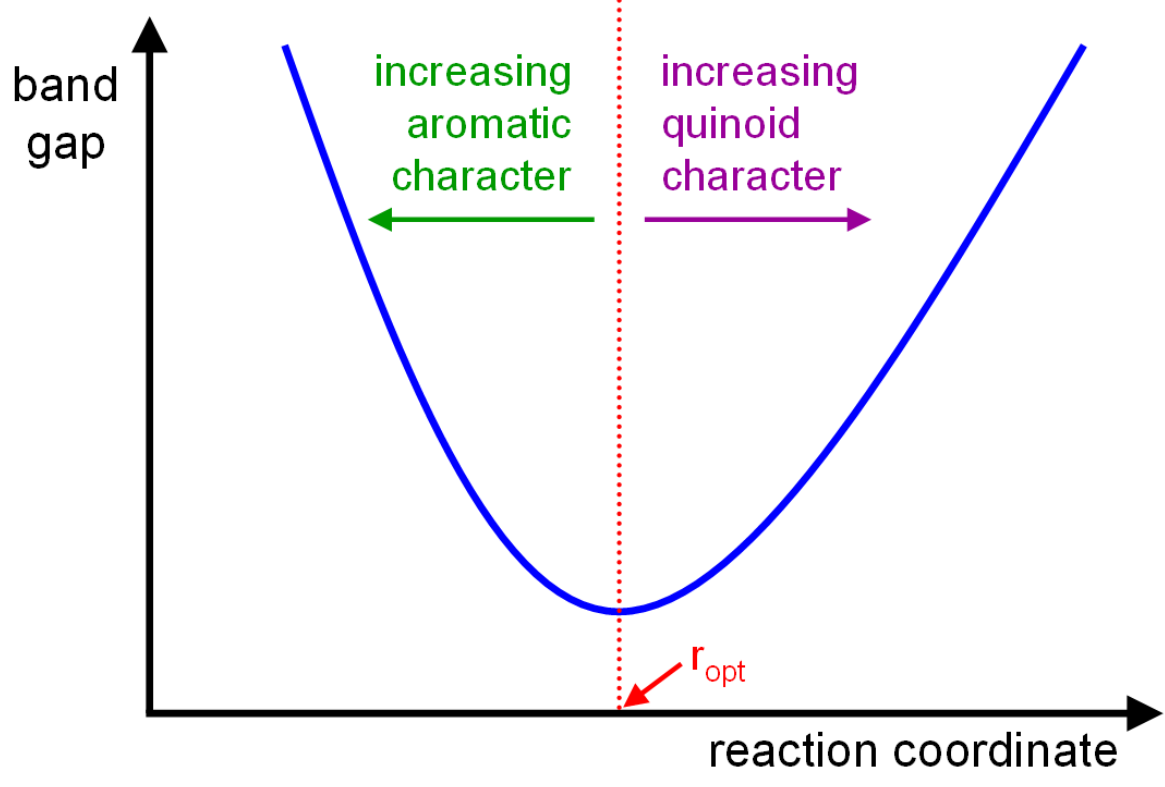

**Figure 7**

**Figure captions**

**Figure 1.** Molecular structures of (a) five-membered heterocyclic polymers and (b) benzannulated heterocyclic polymers. In each of these polymers, vinylene linkage groups connect adjacent monomer units along the backbone chain.

**Figure 2.** Electronic band structures for (a) poly-pyrrole, (b) poly-cyclopentadiene, and their benzannulated versions. The lower blue lines denote valence bands, and the upper red lines represent conduction bands. The dashed horizontal line indicates the position of the Fermi energy in the polymer. All band structures were calculated using a 100 *k*-point mesh obtained from B3LYP/6-311G(d,p) periodic DFT calculations.

**Figure 3.** Cartoon of HOCOs and LUCOs for (a) poly-pyrrole and (b) poly-cyclopentadiene. The ground state of poly-pyrrole is aromatic; destabilization of this state moves the system through a level-crossing to a higher-lying quinoid form, effectively reducing the band gap. In contrast, the ground state of poly-cyclopentadiene is quinoid, and destabilizing its electronic character creates an aromatic form which lies higher in energy (also reducing the band gap).

**Figure 4.** HOCOs and LUCOs depicting the benzannulation of (a) poly-pyrrole and (b) poly-cyclopentadiene. Fusing a benzene ring onto poly-pyrrole destabilizes the HOCO and stabilizes the LUCO, effectively reducing the band gap. However, the same benzannulation procedure stabilizes the HOCO of poly-cyclopentadiene even further and moves the orbitals away from the level-crossing, resulting in a very large band gap.

**Figure 5.** Evolution of band gaps for poly-pyrrole and its derivatives from B3LYP/6-311G(d,p) periodic DFT calculations. The addition of one benzene ring shifts the system closer to an energy level-crossing, resulting in a small band gap; however, the fusion of additional benzene rings moves the system further away from the level-crossing, effectively increasing the band gap (see Figure 7).

**Figure 6.** Correlation between band gaps and bond length alternation (BLA) values for both the five-membered heterocyclic polymers and the benzannulated heterocyclic polymers. In each case, the band gap of the polymer attains its lowest value when the BLA approaches zero.

**Figure 7.** Schematic of band gap energies as a function of increasing aromatic/quinoidal character. The band gap of a polymer attains its lowest value when the reaction coordinate is at the level-crossing denoted by $r_{opt}$ (note the overall similarity with Figure 6). Controlling the relative distance from the level-crossing, $r - r_{opt}$, is key to designing a low-band-gap polymer; being aromatic or quinoidal in itself (represented by the sign of $r - r_{opt}$) does not insure a small band gap.

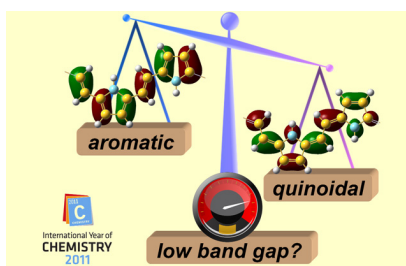